\documentclass[twocolumn]{aastex62}

\usepackage{graphics,epsf}
\usepackage{amsmath}                
\usepackage{amsfonts}               
\usepackage{amssymb}                
\usepackage{epsfig}                 
\usepackage{graphicx}
\usepackage{float}
\usepackage{color}
\usepackage[para,online,flushleft]{threeparttable}

\def \s{~\rm{s}}
\def \km{~\rm{km}}

\def \erg{~\rm{erg}}

\def \yr{~\rm{yr}}

\begin{document}

\title{Minutes-delayed jets from a neutron star companion in core collapse supernovae}



\author[0000-0003-0375-8987]{Noam Soker}
\affiliation{Department of Physics, Technion, Haifa, 3200003, Israel; soker@physics.technion.ac.il}
\affiliation{Guangdong Technion Israel Institute of Technology, Shantou 515069, Guangdong Province, China}


\begin{abstract}
I study cases where a neutron star (NS; or a black hole) companion to a type Ib or type Ic (stripped-envelope) core collapse supernova (CCSN) accretes mass from the explosion ejecta and launches jets minutes to hours after explosion.
The NS orbits at a pre-explosion radius of $a \simeq 1-5 R_\odot$. I find that when the ejecta velocity drops to be $v_{\rm ej} \la 1000-1500 \km \s^{-1}$ the ejecta gas that the NS accretes possesses sufficient specific angular momentum to form an accretion disk around the NS. The NS accretes a fraction of $(M_{\rm acc, d}/ M_{\rm ej}) \approx 3 \times 10^{-5} - 3 \times 10^{-4}$ of the ejecta mass through an accretion disk over a time period of $t_{\rm jets} \approx 10{~\rm min} - {\rm few~hour}$. If the jets carry about ten per cent of the accretion energy, then their total energy is a fraction of about $0.003-0.03$ of the kinetic energy of the ejecta. The implications of these jets from a NS (or a black hole) companion to a CCSN are the shaping of the inner ejecta to have a bipolar morphology, energising the light curve of the CCSN, and in some cases the possible enrichment of the inner ejecta with r-process elements. 
\end{abstract}



\section{Introduction}
\label{sec:intro}
  
There are varieties of scenarios to explain a variety of violent events that involve the spiralling-in of neutron stars (NS) inside the envelope of massive giant stars, i.e., a common envelope evolution (CEE; the same holds for a black hole companion; in most cases when I mention a NS I refer also to a black hole). Some scenarios end with the merger of the NS with the core, while in others the NS survives to the end of the core evolution; the core ends either as a core collapse supernova (CCSN) or in forming a white dwarf. 
 
Scenarios where the NS survives include the formation of close NS binary systems that might later merge by emitting gravitational waves (e.g., \citealt{Taurisetal2017, Kruckowetal2018, MandelFarmer2018, VignaGomezetal2020}), the recycling of a pulsar by accretion of angular momentum during the the CEE (e.g., \citealt{Chattopadhyayetal2020}), and common envelope jets supernova (CEJSN) impostors (e.g., \citealt{Gilkisetal2019a}) where a NS that orbits inside the envelope launches jets (e.g., \citealt{ArmitageLivio2000, Papishetal2015, SokerGilkis2018, LopezCamaraetal2020}).
These cases are likely to end as a binary system of a NS with a stripped-envelope CCSN progenitor, i.e., progenitors of type Ib or type Ic CCSNe (SNe Ibc; e.g., \citealt{Dewietal2002, VignaGomezetal2018, Laplaceetal2020}). 
Scenarios where the NS spirals all the way to merge with the core of the giant star include CEJSN events (e.g., \citealt{Chevalier2012, Sokeretal2019, Schroderetal2020}), including the nucleosynthesis of r-process elements in CEJSNe (e.g., \citealt{GricnenerSoker2019}).   

In some of the cases where the NS (or black hole) survives the CEE, it ends in a binary system with a CCSN  Ibc progenitor with an orbital separation of $a \approx 1-5 R_\odot$, corresponding to an orbital period of $\approx 1-10{~\rm hours}$ (e.g., \citealt{Fragosetal2019, RomeroShawetal2020}). These systems are the target of the present study. I study the accretion of mass by the NS from the ejecta after the CCSN explosion of the core. Cooling by neutrinos allows NS to accrete mass at a high rate when the mass accretion rate is $\dot M_{\rm acc} \ga 10^{-3} \rm M_\odot \yr^{-1}$ \citep{HouckChevalier1991, Chevalier1993, Chevalier2012}. The flow of this process resembles more an accretion from a stellar wind that an accretion inside a common envelope.  
Previous studies of NS accreting from stellar winds include early analytical studies (e.g., \citealt{IllarionovSunyaev1975, ShapiroLightman1976, Wang1981}), more recent analytical studies (e.g., \citealt{DelgadoMarti2001, Erkutetal2019}), and numerical simulations of this accretion process through an accretion disk (e.g., \citealt{ElMellahCasse2017, ElMellahetal2018, XuStone2019}). As well, there are some observational support to the formation of accretion disks via wind accretion (e.g., \citealt{Liaoetal2020}). 

There are early studies of ejecta that collide with a NS companion (e.g., \citealt{EgorovPostnov2009}), and, more relevant to the present study, of a NS accreting mass from a CCSN ejecta (\citealt{Fryeretal2014, Becerraetal2015, Becerraetal2016, Becerraetal2019}). In section \ref{sec:TheFlow}
I describe the flow structure and where I differ from these earlier studies. 
In section \ref{sec:specificJ} I study the conditions for the formation of an accretion disk, in section \ref{sec:MassAccretion} I estimate the  mass that the NS accretes and the energy of the jets it might launch, in section \ref{sec:RelevantCCSNe} I mention the relevant CCSNe for this scenario, and in section \ref{sec:Energy} I compare the energy of the jets to that of the CCSN ejecta. I summarise the results and discuss possible implications of these jets in section \ref{sec:implicaitons}. 
I note that there is another type of jets that might be active minutes after a CCSN explosion, but of a single star. The source of these jets is an accretion disk that an early fallback, within tens of minutes, forms in some CCSNe \citep{Stockingeretal2020}. These jets might have some similar effects to those that I discuss in section \ref{sec:implicaitons} for CCSNe in a close orbit with an NS. 

\section{The flow structure}
\label{sec:TheFlow}

\begin{figure}
\includegraphics[trim=0.6cm 8.8cm 5.1cm 1.9cm ,clip, scale=0.62]{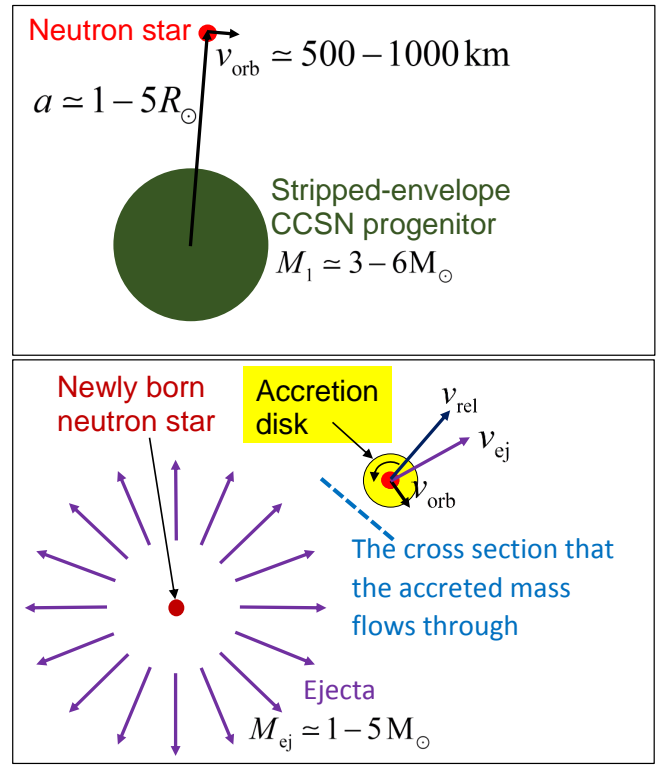}
	\caption{A schematic drawing (not to scaled) in the orbital plane of the binary system before explosion of the stripped-envelope CCSN (upper panel) and during the accretion phase through a disks (lower panel). The NS companion and/or the NS remnant can be black holes. 
	}
	\label{fig:schemetic}
\end{figure}

I consider a NS of mass $M_{\rm NS}$ and radius $R_{\rm NS}$ that orbits the progenitor of a SN Ibc in a circular orbit of radius $a \approx 1 - 5 R_\odot$. I present the pre-explosion system in the upper panel of Fig. \ref{fig:schemetic}. The orbital period is $P_{\rm orb} \approx 1-10 ~{\rm  h}$. I consider the case where the NS accrets mass from the explosion ejecta. Because of the orbital motion, the symmetry axis of the Bondi--Hoyle--Lyttleton (BHL) accretion flow, i.e., the accretion line, is tilted with respect to the ejecta velocity. As a result of that, the density and velocity are not uniform on the cross section (area perpendicular to the accretion line; lower panel of Fig. \ref{fig:schemetic}) of the accretion tube (cylinder), which is the volume through which the gas flows onto the NS. This results in a net angular momentum of the accreted gas, that I calculate in section \ref{sec:specificJ}. \cite{Becerraetal2019} present hydrodynamical simulations of this flow structure.
      
\cite{Becerraetal2015} performed a detail study of this accretion process. I differ from them in the following aspects. 
(1) I use results from three-dimensional hydrodynamical numerical simulations (\citealt{Livioetal1986, Ruffert1999}) to estimate the specific angular momentum of the BHL accretion process, rather than performing a full analytical calculation. 
A full analytical calculation does not capture all aspects of the accretion flow \citep{DaviesPringle1980}. 
(2) \cite{Becerraetal2019} simulated 39 cases, of which one case has an initial orbital radius of $1 R_\odot$ and the rest have orbital radii in the rang of $0.19 - 0.5 R_\odot$. I consider initial orbital radii larger by a factor of about 5, $a \simeq 1- 5 R_\odot$. 
(3) \cite{Becerraetal2015} focused on the spin-up of the NS. My view is that jets remove most of the angular momentum of the gas in the accretion disk. Namely, the accretion process leads to relatively energetic jets. 
(4) My focus is the role that the jets might play in shaping and energising the CCSN ejecta (but I do not refer to gamma ray bursts). 

Later, this group performed hydrodynamical simulations of this accretion flow (\citealt{Becerraetal2016, Becerraetal2018, Becerraetal2019}). They showed that the accretion flow forms an  accretion disk around the NS, but they did not discuss the properties of possible jets and the influence of the jets on the ejecta. These are the focus of the present study.

\section{Accretion disk formation}
\label{sec:specificJ}

I examine the specific angular momentum of the gas that the  NS accretes. In doing so, I follow \cite{Wang1981} who considered an NS that accretes mass from a stellar wind. 
\cite{Becerraetal2015} also followed \cite{Wang1981} in their analytical calculation of a NS accreting mass from a CCSN ejecta. I differ from these two studies in using a factor (the parameter $\eta_a$ below) from numerical simulations of the BHL accretion process.
This factor already incorporates some steps in a pure analytical calculation of the accreted angular momentum, like the shape of the cross section through which the accreted mass flows onto the NS. 

The situation here is more complicated than what \cite{Wang1981} studies, and introduces several uncertainties. (1) The explosion removes mass and the orbit changes, and it is not circular anymore during the accretion phase. The relevant accreted mass crosses the orbital radius in a time of $\tau_c \approx 1 R_\odot /10^3 \km \s^{-1} \approx 0.2 h$, during which the NS moves a large fraction of the orbit. Therefore, a circular orbit is an approximation. (2) The density profile is not of a wind with a constant mass loss rate. (3) I assume a homologous ejecta flow (before the gravity of the NS influences the flow) which is achieved when the Mach number is very large. This is not accurate close to the exploding star (see below). (4) I expect many clumps in the ejecta that introduce stochastic angular momentum variations that change the accretion flow structure(e.g., \citealt{Karino2015} for accretion from a clumpy wind). These might increase the likelihood of the formation of an intermittent accretion disk around the NS.    
  
Despite these uncertainties I follow the derivation of \cite{Wang1981} (and of \citealt{Becerraetal2015}) to some extent.  
I start with the following ejecta density profile long enough after the explosion (here it is about several minutes) such that the velocity at each radius is $v_{\rm ej}(r)=r/t$ (\citealt{SuzukiMaeda2019}; their equation 1-6, with $\delta=1$ and $m=10$), 
\begin{equation}
\rho (r, t) = \begin{cases}
        \rho_0 \left( \frac{r}{t v_{\rm br}} \right)^{-1} 
        & r\leq t v_{\rm br}
        \\
        \rho_0 \left( \frac{r}{t v_{\rm br}} \right)^{-10} 
        & r>t v_{\rm br}, 
        \end{cases}
\label{eq:density_profile}
\end{equation}
    \newline
where $M_{\rm ej}$ is the ejecta mass, $E_{\rm SN}$ is its kinetic energy,  
\begin{eqnarray}
\begin{aligned} 
& v_{\rm br} = \left( \frac{20}{7} \right)^{1/2} \left( \frac {E_{\rm SN}}{M_{\rm ej}} \right)^{1/2} 
=6400
\\& 
\times \left( \frac {E_{\rm SN}}{10^{51} \erg} \right)^{1/2}
\left( \frac {M_{\rm ej}}{3.5 M_\odot} \right)^{-1/2}
\km \s^{-1} ,
\end{aligned}
\label{eq:vbr}
\end{eqnarray}
and
\begin{equation}
\rho_0 = \frac {7 M_{\rm ej}}{18 \pi v^3_{\rm br} t^3} .
\label{eq:rho0}
\end{equation}
I scale the ejecta mass with about the average value of the models of \cite{Dessartetal2016} and \cite{Teffsetal2020}.
This ejecta mass implies a progenitor mass just before explosion of $M_1 \simeq5 M_\odot$ that leaves an NS remnant. This is, for example, also the ejecta mass that \cite{Taddiaetal2019} infer for the SN Ic iPTF15dt. 
For this ejecta mass the two NS remnants are unbound (I do not require them to stay bound). I note that the ejection of most of the mass is on a time scale that is not much shorter than the orbital period. This, together with the mass that the NS accretes, imply, as \cite{Becerraetal2015} already claimed, that the explosion can eject more than $50 \%$ of the initial binary mass and the system still stays bound. In any case, even in a case that the binary stays bound the orbit becomes eccentric and at early times the NS moves away from the explosion center.
    
The mass that the NS accretes through an accretion disk has a low velocity (see below), for which the density profile has a power law of $-1$. The density and velocity derivatives with radius at a given time are therefore
\begin{equation}
\frac{\rho^\prime}{\rho} = - \frac{1}{r} ; \qquad  \frac{v_{\rm ej}^\prime}{v{\rm ej}} = \frac{1}{r}  .
\label{eq:rhoVprime}
\end{equation}

In his equation (17)  \cite{Wang1981} gives the angular momentum inflow rate though an area $dy dz$ on the cross section of the accretion tube under the assumption that the wind mass loss rate is constant. I repeat that derivation but with the density and velocity derivatives from equation (\ref{eq:rhoVprime}). With the aid of equation (\ref{eq:rhoVprime}) at $r=a$ I replace equation (17) of \cite{Wang1981} with the following expression for the rate of angular momentum flow through an area $dy dz$ 
\begin{eqnarray}
\begin{aligned} 
dN & \simeq \rho(a) v^2_{\rm rel} (a) 
\left\{ 1+ \left[-\frac{\rho^\prime}{\rho} -2 \frac{v_{\rm ej}^\prime}{v_{\rm ej}} \right] y \sin \alpha_0 \right\} y dy dz 
\\& 
\simeq  \rho(a) v^2_{\rm rel} (a) 
\left\{ 1 - \frac{1}{a} y \sin \alpha_0 \right\} y dy dz
,
\end{aligned}
\label{eq:dN}
\end{eqnarray}
where $\sin \alpha_0 \equiv v_{\rm orb}/v_{\rm rel}$, $v_{\rm orb}$ is the relative orbital velocity of the NS and the CCSN progenitor, and 
 \begin{equation}
 v_{\rm rel} \simeq [v^2_{\rm ej}(r) + v^2_{\rm orb}]^{1/2}, 
 \label{eq:Vrel}
\end{equation}
is the relative ejecta-NS velocity. The expression for the relative velocity is an approximation here (unlike in the case that \citealt{Wang1981} studies) because the orbit is not circular anymore.
Therefore, the pre-explosion orbital velocity is not the accurate velocity, and the two velocities are not perpendicular to each other. Like \cite{Becerraetal2015}, I will stay with this approximation.   
 
The derivatives of the density and relative velocity (equation \ref{eq:rhoVprime}), which represents the gradient of the density and velocity on the cross section of the accretion tube (cylinder), contribute with opposite signs to the net angular momentum in equation (\ref{eq:dN}). For a wind with a constant mass loss rate the density gradient dominates and the sense of angular momentum of the accreted mass is the same as that of the binary system (e.g.,  \citealt{ShapiroLightman1976}). However, in the present case the density decrease is shallower, $\rho \propto r ^{-1}$ instead of $\rho \propto r ^{-2}$, and the ejecta velocity increases outward as $v_{\rm ej} \propto r^{1}$ at a given time. 
The outcome is that the sense of the angular momentum of the accreted mass is opposite to that of the binary system (see Fig.2 of \citealt{Becerraetal2019}). 

The absolute value of the second term in the last line of equation (\ref{eq:dN}) is like that in a wind with a constant mass loss rate and a velocity of $v_{\rm w} \propto r^{1}$, but as mentioned above the sense of angular momentum is opposite. There is an uncertainty of the boundary of the cross section over which we should carry the integration (e.g., \citealt{DaviesPringle1980}).
To overcome this uncertainty I adopt the approach (e.g., \citealt{SokerRappaport2000}) of using numerical results to scale the accreted angular momentum that \cite{Wang1981} derives (in doing so I deviate from the calculation of \citealt{ Becerraetal2015}). These numerical results (e.g., \citealt{Livioetal1986, Ruffert1999}) show that the specific angular momentum of the accreted mass is only a fraction of $\eta_a \simeq 0.2$ of what the simple BHL accretion gives. 
 Over all, the specific angular momentum of the accreted mass is as the value that \cite{Wang1981} derives multiplied by $\eta_a$, 
\begin{equation}
j_a \approx \frac{1}{2} \eta_a 
\left( \frac {2 \pi } {P_{\rm orb}} \right) R^2_a ,
\label{eq:jacc}
\end{equation}
where 
\begin{equation}
R_a = \frac{2 G M_{\rm NS}} {v^2_{\rm rel}} =  0.16  
\left( \frac {M_{\rm NS}}{1.4 M_\odot} \right)
\left( \frac {v_{\rm rel}}{1800 \km \s^{-1}} \right)^{-2}  R_\odot, 
\label{eq:Ra}
\end{equation}
\newline
is the accretion radius. For the accretion rate I take the BHL accretion radius because numerical simulations show that the actual accretion radius is only about $10 \%$ smaller (e.g., \citealt{Ohsugi2018}).

The condition for the formation of an accretion disk is $j_a > j_{\rm NS}$, where 
\begin{equation}
j_{\rm NS} = 
\left( G M_{\rm NS} R_{\rm NS} \right)^{1/2}
\left(1 - \frac {3 G M_{\rm NS}}{c^2 R_{\rm NS}} \right)^{-1/2} 
\label{eq:Jns}
\end{equation}
is the specific angular momentum of a particle in a circular orbit at the equator of the accreting NS. 
For $M_{\rm NS}=1.4 M_\odot$ and $R_{\rm Ns} =12 \km$ the second parenthesis in equation (\ref{eq:Jns}) contributes a factor of $1.44$, which I will use throughout. Substituting typical values for the present study gives the condition for accretion disk formation \citep{SokerRappaport2000}
\begin{eqnarray}
\begin{aligned} 
& 1< \frac{j_a}{j_{\rm NS}} \approx   
\left( \frac {\eta_a}{0.2} \right)
\left( \frac {M_1+M_{\rm NS}}{6 M_\odot} \right)^{1/2}
\left( \frac {M_{\rm NS}}{1.4 M_\odot} \right)^{3/2}
\\& 
\left( \frac {R_{\rm NS}}{12 \km} \right)^{-1/2}
\left( \frac {a}{1 R_\odot} \right)^{-3/2}
\left( \frac {v_{\rm rel}}{1800 \km \s^{-1}} \right)^{-4} , 
\end{aligned}
\label{eq:jajNS}
\end{eqnarray}
where $M_1$ is the supernova progenitors mass just before explosion. 
This expression that is based on the parameter $\eta_a$ that I take from numerical simulations, differs in its usage of parameters from the condition that \cite{Becerraetal2015} derive, although the content is the same.

I scale the orbital separation in equation (\ref{eq:jajNS}) with the lower boundary of the  $a \approx 1-10 R_\odot$ range of values that observations and models give for most systems (e.g., \citealt{Kruckowetal2018}). An NS in an orbit with a periastron distance of $a_p \simeq 1 R_\odot$ requires the primary star to be stripped from all its hydrogen and most of its helium. Progenitors of SNe Ib that have a substantial helium layer but no hydrogen and with a pre-explosion mass of a $M_1 \simeq 4-7 M_\odot$ have radii of $R_1 \simeq 1-2 R_\odot$ (e.g., \citealt{Gilkisetal2019b}). In that case the relevant orbital radius (for a circular orbit) is $a \simeq 3-10 R_\odot$. 
      
As I indicated in section \ref{sec:specificJ}, equation (\ref{eq:jajNS}) is an approximate expression. I actually expect the accreted mass to have more temporarily specific angular momentum due to clumps in the inner ejecta. 
As well, equation (\ref{eq:density_profile}) assumes a homologous expansion far from the exploding star, which is not accurate as the accreting NS is close to the exploding star. The actual velocity is lower than the final one because the thermal pressure is not negligible at early phases of the expansion (before the ejecta suffers substantial adiabatic cooling). In any case, as $a>R_1$, the thermal pressure is smaller than the kinetic energy. 
If we to include the  thermal pressure, then we would replace $v^2_{\rm rel}$ in the denominator of equation (\ref{eq:Ra}) for the accretion radius with $v^2_{\rm rel,P} + C^2_{\rm s}$, where $C_{\rm s}$ is the sound speed in the ejecta and $v_{\rm rel,P}$ is the ejecta velocity when pressure is still important (i.e., $C^2_{\rm s}$ is not much smaller than $v^2_{\rm rel,P}$). From energy conservation we expect that
$v^2_{\rm rel} \simeq v^2_{\rm rel,P} + C^2_{\rm s}$, so that the accretion radius does not change much. 
As for the condition for an accretion disk formation, since the ejecta velocity is smaller the influence of the orbital motion, that breaks the symmetrical accretion flow, is larger. This results in a larger specific angular momentum of the accreted mass, allowing a larger relative velocity $v_{\rm rel}$, therefore a larger ejecta velocity, for disk formation.
Namely, accretion through a disk starts earlier, and therefore the NS accretes more mass through a disk. 
I conclude that the assumption of a homologous ejecta expansion is adequate for the present goals. 
  
Staying with equation (\ref{eq:jajNS}) and its scaling, the condition for accretion disk formation is $v_{\rm rel} \la 1800 \km \s^{-1}$. By equation (\ref{eq:Vrel}) it becomes a condition on the ejecta velocity $v_{\rm ej} \simeq (v^2_{\rm rel} - v^2_{\rm orb})^{1/2}$. 
For the parameters of equation (\ref{eq:jajNS}) the relative orbital velocity of the two stars is $v_{\rm orb} \simeq 1000 \km \s^{-1}$, and so the condition for accretion disk formation becomes 
\begin{eqnarray}
\begin{aligned}
& v_{\rm ej} <  v_{\rm ej,d} \approx 1500  
\bigg[   \frac{3.24}{2.24} 
\left(  \frac{v_{\rm rel}}{1800 \km \s^{-1}}  \right)^2 
\\ & - \frac{1}{2.24}  \left(  \frac{v_{\rm orb}}{1000 \km \s^{-1}}  \right)^2 
\bigg]^{1/2} 
\km \s^{-1}. 
\end{aligned}
\label{eq:Vejd}
\end{eqnarray}
The above value is about quarter of $v_{\rm br}$ (eq. \ref{eq:vbr}), 
$v_{\rm ej,d} \approx 0.235 v_{\rm br}$. 

For a circular orbit with $a=5$, keeping the other parameters as in the above equations, the orbital velocity is $v_{\rm orb} \simeq 500  \km \s^{-1}$, 
equation (\ref{eq:jajNS}) gives the limit on the relative velocity as $v_{\rm rel} \la 1000 \km \s^{-1}$, and from equation (\ref{eq:Vejd}) the maximum velocity of the ejecta to form an accretion disk is $v_{\rm ej,d} \approx 900 \km \s^{-1}$.

\section{Mass accretion and jets' energy}
\label{sec:MassAccretion}

I turn to estimate the mass that is accreted through an accretion disk.
Equation (\ref{eq:Vejd}) constrains the ejecta velocity to $v_{\rm ej} < v_{\rm ej,d}$. From equation (\ref{eq:density_profile}) the total amount of ejecta mass in that range is
\begin{eqnarray}
\begin{aligned}
& M_{\rm ej} (<v_{\rm ej,d}) = \int^{v_{\rm ej,d} t}_0 \rho_0 \left( \frac{r}{t v_{\rm br}} \right)^{-1} 4 \pi r^2 dr 
\\ &
= \frac{7}{9} \left( \frac{v_{\rm ej,d}} {v_{\rm br}} \right)^2 M_{\rm ej} = 0.15 
\left( \frac{v_{\rm ej,d}}{0.235 v_{\rm br}} \right)^2 
\left( \frac{M_{\rm ej}}{3.5 M_\odot} \right) M_\odot. 
\end{aligned}
\label{eq:Mej(Vejd)}
\end{eqnarray}
The NS companion accretes only a small fraction of this mass, that I estimate from the BHL accretion flow by using the accretion radius as in equation (\ref{eq:Ra}).
Crudely, the mass that the NS accretes via an accretion disk is 
\begin{eqnarray}
\begin{aligned}
 M_{\rm acc,d} & \approx 
M_{\rm ej} (<v_{\rm ej,d}) \frac { \pi R^2_a}{4 \pi a^2} 
\\ &
\approx 10^{-3}  
\left( \frac{ M_{\rm ej} (<v_{\rm ej,d}) }{0.15M_\odot} \right)
\left( \frac {M_{\rm NS}}{1.4 M_\odot} \right)^2
\\ &  \times
\left( \frac {v_{\rm rel}}{1800 \km \s^{-1}} \right)^{-4} 
\left( \frac{a}{1 R_\odot} \right)^{-2} M_\odot 
 . 
\end{aligned}
\label{eq:Macc(Vejd)}
\end{eqnarray}
  
I note the following. ($i$) The accretion radius increases with time as the ejecta velocity decreases. This increases the accretion rate. ($ii$) The NS acquires a post-explosion eccentric orbit if it is still bound to the newly born NS from the  explosion, and a hyperbolic orbit if the binary system becomes unbound. For that, during the accretion phase the distance increases from its pre-explosion value. This reduces the accretion rate. ($iii$) I expect that polar jets drive the explosion because the close NS companion spins-up the CCSN progenitor. Such an explosion in rapidly rotating cores would leave a slower equatorial outflow compared to spherical explosions \citep{Gilkisetal2016Super}.
The slower velocities of the equatorial ejecta means that both more mass has sufficient angular momentum to form an accretion disk (equation \ref{eq:jajNS}), and a larger accretion radius (equation \ref{eq:Ra}).  
Overall, there are large uncertainties, starting from the uncertainties in equation (\ref{eq:jajNS}) for the condition on disk formation, and continuing with the uncertainties in the value of the accretion radius, both of which contribute to large uncertainties in the value that equation (\ref{eq:Macc(Vejd)}) gives.  

The accretion process through a disk stars at $\simeq a/v_{\rm ej,d} \simeq 8{~\rm min} - 1 ~{\rm hour}$ for the scaling of this study ($a \simeq 1-5 R_\odot$),  and might be significant until the NS moves away $\simeq {\rm few} \times a/ v_{\rm orb} \approx 1 - {\rm several} ~{\rm hour}$. 
Because the accretion rate is much above $10^{-3} M_\odot \yr^{-1}$, the accreted mass cools by neutrino emission (\citealt{Chevalier1993}), and the Eddington limit is not relevant. 
Taking that the accretion disk launches jets that carry a canonical fraction of $\zeta \simeq 0.1$ of the accretion energy, e.g., about $10 \%$ of the accreted mass at about the escape velocity, implies that the two jets carry an energy of (most of the rest of the accretion energy is carried by neutrinos)  
\begin{eqnarray}
\begin{aligned}
& 
E_{\rm 2j} \simeq 3 \times 10^{49}
\left( \frac {\zeta}{0.1} \right) 
\left( \frac {M_{\rm acc,d}}{10^{-3} M_\odot} \right) 
\\ &
\times
\left( \frac {M_{\rm NS}}{1.4 M_\odot} \right)
\left( \frac {R_{\rm NS}}{12 \km} \right)^{-1} \erg  . 
\end{aligned}
\label{eq:E2j}
\end{eqnarray}
  
For a circular orbit with $a=5 R_\odot$ (keeping all other parameters the same as in the above equations), for which the conditions for disk formation is $v_{\rm rel} \la 1000 \km \s^{-1}$ and so $v_{\rm ej,d} \approx 900 \km \s^{-1}$ (section \ref{sec:specificJ}), equations (\ref{eq:Mej(Vejd)}), (\ref{eq:Macc(Vejd)}) and (\ref{eq:E2j}) give $M_{\rm ej}(<v_{\rm ej,d}) \approx 0.05 M_\odot$, $M_{\rm acc,d} \approx 10^{-4} M_\odot$, and $E_{\rm 2j} \approx 3 \times 10^{48} \erg$, respectively. 

\section{Types of relevant CCSNe}
\label{sec:RelevantCCSNe}

We should distinguish between cases with SNe Ib and SNe Ic. The progenitors of SNe Ic have a typical pre-explosion radius of $R_1  < 1 R_\odot$. This allows a NS companion to orbit at $a \approx 1R_\odot$. Progenitors of SNe Ib have helium layer, and so their typical radius is larger. I scale here the ejecta mass with $M_{\rm ej} = 3.5 M_\odot$, both for SNe Ib and SNe Ic (e.g., \citealt{Teffsetal2020}). 
 
It is possible that in the presence of a close NS companion the explosion will be of the ultra-stripped type (e.g., \citealt{Taurisetal2015}), ejecting only $M_{\rm ej} \approx 0.05-0.2 M_\odot$ \citep{Taurisetal2013, Hijikawaetal2019}. 
Even in some of these cases the jets from a NS comapnion might be significant. Consider the Ca-rich SN~2019ejh. \cite{JacobsonGalanetal2020} propose that this SN was a peculiar SN Ia in the frame of the double-degenerate scenario. \cite{Nakaokaetal2020}, on the other hand, propose that SN~2019ejh was an ultra-stripped envelope CCSN, and estimate the ejecta mass and kinetic energy to be $M_{\rm ej} \simeq 0.43 M_\odot$, and $E_{\rm SN} \simeq 1.7 \times 10^{50} \erg$, respectively. I consider the later model. From equation (\ref{eq:vbr}) the velocity at the  break of the density profile is $v_{\rm br} \simeq 7500 \km \s^{-1}$, that is not much different than the scaled value I use in this study ($6400 \km \s^{-1}$). For the same orbit, the orbital velocity is somewhat smaller due to lower SN progenitor mass. This reduces the relative jet-NS velocity, which it turns increases the accretion radius. Overall, the NS companion in such a low-mass and low-energy CCSN might accreted a similar fraction of the ejecta mass through an accretion disk as what the equations in section \ref{sec:MassAccretion} give. The implications of jets from a NS companion in such an ultra-stripped envelope CCSN are similar to what I discuss below for CCSNe Ibc with more massive envelopes. 

\section{Energy Considerations}
\label{sec:Energy}

Equation (\ref{eq:E2j}) shows that the jets' energy is much smaller than the explosion energy. However, it is not negligible with respect to the kinetic energy of the inner part of the ejecta from which the NS accretes mass through a disk $E_{\rm ej} (<v_{\rm ej,d})$, and with respect to the energy in radiation $E_{\rm rad} \approx 10^{49} \erg$.
The total kinetic energy of the mass  $M_{\rm ej} (<v_{\rm ej,d})$ is 
\begin{eqnarray}
\begin{aligned}
& E_{\rm ej} (<v_{\rm ej,d}) = \int^{v_{\rm ej,d} t}_0 \rho_0 \left( \frac{r}{t v_{\rm br}} \right)^{-1} \frac{1}{2} v^2 4 \pi r^2 dr 
\\ &
= \frac{5}{9} E_{\rm SN}  
\left( \frac{v_{\rm ej,d}} {v_{\rm br}} \right)^4  
 = 1.7 \times 10^{-3}    
\left( \frac{v_{\rm ej,d}}{0.235 v_{\rm br}} \right)^4 E_{\rm SN}.
\end{aligned}
\label{eq:Eej(Vejd)}
\end{eqnarray}
   
For an explosion energy of $E_{\rm SN} = 10^{51} \erg$ the jets carry an order of magnitude more energy than the energy of the slow ejecta it accretes from through a disk. For the scaling I use here, the energy of the jets (eq. \ref{eq:E2j}) is equal to the kinetic energy of ejecta mass slower than a velocity of about $3000 \km \s^{-1}$, which amounts to a mass of $\simeq 0.18 M_{\rm ej}$. 
However, the flow structure is far from being spherically symmetric. The jets interact mainly with the polar ejecta. For $E_{\rm 2j} =0.03 E_{\rm SN}$, 
the jets carry an energy equals to that of the ejecta outflowing within two opposite polar cones with a half opening angle of $14^\circ$. Namely, narrow jets can penetrate deep into the ejecta. 

\section{Summary and implications}
\label{sec:implicaitons}

I considered a stripped-envelope CCSN in the presence of an NS companion at an orbital separation of $a \simeq 1-5 R_\odot$. I estimated that when slower ejecta flows around this NS, $v_{\rm ej} \la 1000-1500 \km \s^{-1}$, the NS accretes mass through an accretion disk that launches jets (equations \ref{eq:jajNS} and \ref{eq:Vejd}). Without an accretion disk, the NS will not launch jets, and I expect neutrinos to carry close to 100\% of the accretion energy. The accretion disk is very likely to launch jets that carry $\approx 10\%$ of the accretion energy. The jets' active phase lasts for a period of $t_{\rm jets} \approx 10{~\rm min} - 1{~\rm hour}$ for $ a \simeq 1 R_\odot$ and $t_{\rm jets} \approx 1{~\rm hour}1 - {~\rm several~hour}$ for $ a \simeq 5 R_\odot$ after explosion, the same as the accretion through a disk phase. 

I estimated that the NS accretes a fraction of $(M_{\rm acc, d}/ M_{\rm ej}) \approx 3 \times 10^{-5} - 3 \times 10^{-4}$ of the ejecta mass through an accretion disk. If the jets carry about 10\% of the accreted mass at the escape velocity from the NS, the energy they carry (equation \ref{eq:E2j}) is non-negligible with respect to, or larger even than, the energy of the inner parts of the ejecta (equation \ref{eq:Eej(Vejd)}). As well, the energy of the jets is comparable to the energy of the ejecta within an angle of $\simeq 10-20^\circ$ from the symmetry axis (jets' axis).  

Below I list the possible effects that the jets that a close NS companion launches might have on the ejecta. But first I note that with the appropriate scaling, all calculations and implications of this study are relevant to a black hole companion to a CCSN. 

In a recent thorough numerical study of CCSN explosions (in single stars), \cite{Stockingeretal2020} find that in some cases there is an early fall back of material, over minutes to hours, that has sufficient specific angular momentum to form an accretion disk around the newly born NS. The mass of the disk in their best case for disk formation, out of three cases, is similar to the mass I find in equation (\ref{eq:Macc(Vejd)}), and the time scales after explosion are similar. I therefore expect these jets to have some similar  effects to those from a NS companion that I list below. 

\subsection{Shaping the inner ejecta to a bipolar structure}
\label{subsec:Bipolar}

The energy of the jets that is more than about 1 percent of the total kinetic energy of the ejecta $E_{\rm 2j} \ga 0.01 E_{\rm SN}$ (eq. \ref{eq:E2j}), and their propagation along the polar direction where they interact with a small fraction of the ejecta, imply that the jets shape the inner ejecta to possess a bipolar structure. I expect the final ejecta morphology to have a more spherical outer regions and bipolar inner regions, similar to those of many planetary nebulae that have an outer spherical halo and an inner bipolar or elliptical morphology (e.g., \citealt{Corradietal2003}). 
We present first numerical simulations of this type of interaction of jets with the inner ejecta in a follow-up paper \citep{AkashiSoker2020}. 
{{{ The jets influence the shaping of ejecta that is slower than $\approx 3500 \km \s^{-1}$. An observational indication might be large polarisation when the photosphere recedes to that mass coordinate.   }}}

\subsection{Light curve}
\label{subsec:LightCurve}

A second possible effect is on the light curve. Although  the jets' energy is much smaller than the ejecta energy, it is of the same order of magnitude as the total radiated energy from a typical CCSN. 
 
The progenitor is a small star $R_1 < a \approx 1R_\odot$, that suffers substantial adiabatic cooling at the first few hours of the explosion. The jets that the NS companion launches during the jets' active phase, $t_{\rm jets} \approx 10 ~{\rm min} - {\rm few} ~{\rm hour}$, interact with the ejecta by shock waves that convert kinetic energy to thermal energy at much larger distances. Therefore, although the extra energy is not large relative to the kinetic energy of the ejecta, the two opposite hot bubbles that the jets form (one `cocoon' for each jet) away from the center suffer much less adiabatic losses at a given time (e.g., \citealt{KaplanSoker2020}). More than that, if the jets are narrow they can penetrate deep into the ejecta and form the two hot bubbles closer to the photosphere. This implies a shorter photon diffusion time out from the ejecta, which in turn converts a larger fraction of the thermal energy to radiation (e.g., \citealt{KaplanSoker2020}).  

In a very recent study \cite{Gaoetal2020} explore a similar effect due to a black hole companion to a CCSN. They study the effect of the jets that the black hole launches as it accrete mass from the ejecta on the light curve.  

\subsection{Nucleosynthesis inside the jets}
\label{subsec:nucleosynthesis}

Another effect, a more speculative one, is that nucleosynthesis of neutron-rich material takes place inside the jets. The density of the accretion disk around the NS is very high and electron capture onto nuclei and protons form neutron-rich material. The launching of such neutron-rich material into the jets might lead to nucleosynthesis of r-process elements in the jets \citep{Fryeretal2014, Becerraetal2016, Becerraetal2019}, much as in CEJSNe (e.g.,  \citealt{GrichenerSoker2019}).

\section*{Acknowledgments}
I thank the anonymous referee and Avishai Gilkis for helpful comments. 
This research was supported by a grant from the Israel Science Foundation (420/16 and 769/20) and a grant from the Asher Space Research Fund at the Technion.


\end{document}